\documentstyle[12pt]{article}

\newcommand{\BEQ}{\begin{equation}}    
\newcommand{\BEA}{\begin{eqnarray}}
\newcommand{\EEQ}{\end{equation}} 
\newcommand{\EEA}{\end{eqnarray}}
\newcommand{\rar}{\rightarrow}    
\newcommand{\rcd}[1]{\bar{\cal D}_{#1}}          
\newcommand{\zeile}[1]{\vskip #1 \baselineskip}  

\catcode`\@=11
\def\numberbysection{\@addtoreset{equation}{section}
        \def\theequation{\thesection.\arabic{equation}}}
\numberbysection

\begin{document}
%
%
\begin{titlepage}
\null
\vskip 1cm
\begin{center}
\vskip 0.5in
{\Large \bf Intermittency studies in directed bond percolation}
\vskip 0.5in
 Malte Henkel$^{a}$\footnote{Supported in part by the Swiss National
Science Foundation} and Robert Peschanski$^{b}$
 \\[.3in]
{\em $^{a}$D\'epartement de Physique Th\'{e}orique,
     Universit\'e de Gen\`eve \\
     24  quai Ernest Ansermet,
     CH - 1211 Gen\`eve 4, Switzerland}
\zeile{1}
{\em $^{b}$ Service de Physique Th\'eorique\footnote{
Laboratoire de la Direction des Sciences de la Mati\`{e}re du
Commissariat \`{a} l'Energie Atomique}, CEN Saclay \\
     F- 91191 Gif sur Yvette Cedex, France}
\zeile{2}
{\bf UGVA-DPT 1992/06-769} \\
{\bf Saclay SPhT/92-069}
\end{center}
%
%
{\bf Abstract} \zeile{1}
The self-similar cluster fluctuations of directed bond percolation at
the percolation threshold are studied using techniques
borrowed from inter\-mit\-ten\-cy-related analysis in multi-particle
production. Numerical simulations based on the factorial moments for
large $1+1$-dimensional lattices allow to handle statistical
and boundary effects
and show the existence of weak but definite intermittency patterns.
The extracted fractal dimensions are in agreement with scaling
arguments leading to a new relation linking
the intermittency indices to the critical exponents and the
fractal dimension of directed percolation clusters.

\end{titlepage}

\newpage
%
%

\section{Introduction}

The transition between integrable and chaotic behaviour has attracted
interest for quite a long time. In particular, in the context of hydrodynamics,
the spatio-temporal intermittent behaviour has been studied \cite{Pom86}. On
the
other hand, similar structures have shown up in the context of
particle fragmentation at high energies, as first exemplified by the
JACEE cosmic-ray event and later on by many studies at particle accelerators
\cite{Bia91}. Intermittency studies have also been performed
in Statistical Mechanics, on systems like the $2D$ Ising model at the
second-order transition \cite{Wos88,Sat89}.
In this work, we consider directed bond percolation as
a simple toy model for intermittent behaviour. As we shall argue below,
it might provide a convenient model approach to describe some basic
properties of intermittent systems.

Our interest in studying intermittency was triggered by the observation
that directed percolation can be used as a simple non-perturbative
model for particle fragmentation as described by the Schwinger
mechanism \cite{Bia89,Bar87}. Already before that, it had been proposed that
directed percolation might be useful in describing the transition
between laminar and turbulent flow, occuring for example in
hydrodynamics \cite{Pom86}. We shall describe these heuristic arguments
in section~2. For the quantitative description of intermittent behaviour,
a practical tool are the factorial moments \cite{Bia86}
\BEQ
F_{q} := \frac{<n(n-1)\cdots (n-q+1)>}{<n>^{q}}
\EEQ
where $n$ is the number of particles in a certain (rapidity) interval
and $<>$ denotes the average over a given sample. We shall recall
some of their properties in section~2. Intermittent behaviour
means that the $F_{q}$ obey a power law as a function of the binning
size $\delta$
\BEQ
F_{q}(\delta) \sim \delta^{-\phi_q}
\EEQ
which defines the intermittency indices $\phi_q$. The verification of
this relation and the eventual calculation of these numbers will be the
first objective of this paper, to be described in section~3.
{}From the $\phi_q$, one obtains the
Renyi codimension $\bar{\cal D}_q$
\BEQ
\bar{\cal D}_q = \phi_q  / (q-1)
\EEQ
If the fluctuations of the model have a fractal nature, the $\bar{\cal D}_q$
will be $q$- independent, while for a multifractal, the $\bar{\cal D}_q$
retain a dependence on $q$. In the case of a simple fractal behaviour,
when $\bar{\cal D}_q = \bar{\cal D}$ for all $q$, we shall obtain a
relationship
with the fractal codimension $\bar{D}$ of the directed percolation cluster
and critical exponents (see section~4)
\BEQ
\bar{\cal D} = \beta /\nu_{\perp} - \bar{D} \geq  0
\EEQ
This result allows to reconsider the recent comparison between numerical
calculations of fractal cluster dimensions with theoretical ideas
\cite{Hed91}. It also casts a bridge with the Reggeon Field Theory, which
furnishes a continuum field theory description of directed percolation
\cite{Car80}.
The validity of the result (1.4) only depends on simple
scaling arguments and should therefore not be restricted to
critical directed percolation systems. Section~5 presents our conclusions.

\section{Directed percolation and intermittency}

\subsection{Some heuristics}

We recall the definition of directed percolation (see e.g. \cite{Kin83}).
Consider a lattice
and single out a preferred direction. This preferred direction will
be called "time", the other ones are referred to as "space".
For the purposes of this paper, we shall limit ourselves exclusively
to the case of one temporal and one spatial dimension.
The sites of the lattice can be either filled $(1)$ with probability
$\tilde{p}$ or empty $(0)$ with probability $1-\tilde{p}$. A bond between sites
is present with probability $p$ and absent with probability
$1-p$. The occupation $v_{i,t}$ of the sites is defined in terms of the
conditional probabilities $P(v_{i,t+1}|v_{i-1,t},v_{i+1,t})$, see fig.~1,
namely,
\BEQ
P(1|0,0)=0 \;\; , \;\; P(1|1,1)=p(2-p)\tilde{p} \;\; ,
\;\; P(1|0,1)=P(1|1,0)=p\tilde{p}
\EEQ
together with $P(0|v,v')=1-P(1|v,v')$.
For $\tilde{p}=1$, one has directed bond percolation
and for $p=1$ one has directed site percolation. The probability ${\cal P}$
for having an infinite directed cluster of connected sites scales
with the distance from the percolation threshold $p_c$ as
${\cal P}\sim (p-p_{c})^{2\beta}$ and has the role of an order parameter.
The scaling of the temporal and spatial
correlation lengths $\xi_{\|}$ and $\xi_{\perp}$ is described by
exponents $\nu_{\|}$ and $\nu_{\perp}$, viz.
$$\xi_{\|} \sim (p-p_{c})^{-\nu_{\|}} \;\; , \;\;
\xi_{\perp} \sim (p-p_{c})^{-\nu_{\perp}}$$
Numerical values are \cite{Ess88}
\BEA
\tilde{p}_c = 0.705489(4) &\;\;& ~\mbox{\rm square site} \nonumber \\
p_c = 0.644701(1) &\;\;& ~\mbox{\rm square bond}
\EEA
and for the exponents \cite{Ess88}
\BEQ
\beta = 0.277(2) \; , \; \nu_{\|} = 1.7339(3) \; , \; \nu_{\perp} = 1.0969(3)
\EEQ
{}From these, all other exponents can be found by scaling relations. For
a test of universality for the exponents and the two-point
correlation scaling function, see \cite{Ben84}.

The problem of directed percolation can be rewritten in terms of
a field theory. As it is well
known, directed bond percolation is in the same universality class as
Reggeon field theory \cite{Car80}, at least close to the upper critical
dimension $D^{*}=5$. Nevertheless, this universality was also confirmed
numerically for $D=2$ and $D=3$, see refs. \cite{Bro78,Hen90,Gra89}.
Consequently, results of a directed percolation study also carry an implication
about that theory. We will discuss that point later on.

Some time ago, Pomeau \cite{Pom86} has proposed that spatial-temporal
intermittency were in the same universality class as directed percolation.
According to him, the transition to the turbulent state in a liquid
is similar to the one induced by the intermittent transition in
oscillations and that the interaction between neighboring
oscillators might be seen as some sort of local contamination process.
It appears then plausible that an oscillator in a turbulent state
may either relax spontaneously towards its quiescent state or else
may contaminate its neighbors. As an example of this, one might
consider the observation that the laminar convection
rolls in Taylor-Couette flow do not break up spontaneously into
a turbulent state but get only
infected via a turbulent neighbor.
This is nothing else than the familar definition of directed percolation.
Although this argument is purely qualitative and does not yield
a value for the percolation probability $p$, it does carry some
heuristic value, even if the simple directed percolation is
probably not sufficient to describe intermittency occuring in realistic
liquids \cite{Man88}.

To complement this, we now argue that directed percolation might be
useful to study some aspects of particle fragmentation.
For simplicity, consider the example of
$e^{+} e^{-}$ annihilation producing a quark-antiquark pair. It is
generally believed that for a confining theory like QCD the color
lines of force are essentially concentrated in a narrow tube connecting
$q$ with $\bar{q}$, acting like a string with a constant tension.
By the Schwinger mechanism, new $q \bar{q}$ pairs are created out of the
vacuum by the color force with a probability $P$ expressible in terms
of known quantities \cite{Bia89,Bar87}. One thus obtains a resulting color
field
which will strongly fluctuate as a function of the relative rapidity
between the particles.

We now make the link from directed percolation to this model.
Since the color flux tube is essentially one-dimensional, we take
a two-dimensional lattice. The direction perpendicular ("space") to the
imposed propagation ("time") direction will be taken to correspond to the
rapidity interval.
On each site of the spatial lattice we put $0$ if there
is no color field and $1$ if there is one. The quarks or antiquarks
generated are placed on the dual spatial lattice.
The creation of a new $q\bar{q}$ pair generates a supplementary color
field which exactly compensates the one which produced the new pair.
In the particle language, this means that some local field energy
has been transformed in particle creation with probability $P$. In
the percolation language, this corresponds to having a site occupied by
$0$ at time $t$ after its two neighbors at time $t-1$ have been
occupied by one. In the frame of directed bond percolation, see (2.1),
we have $P=1-P(1|1,1)=(1-p)^2$. Again, this very
simple model should be taken as a guide but not as a realistic description
of high-energy processes.

The breaking of the color field is stochastic and, due to the causality
nature of directed percolation, new particles can only be created where the
field persits, it generates a cascade. Since randomness and cascading
are just the two features known to give intermittency in particle
production \cite{Bia89}, and because of the connection with turbulent
behaviour \cite{Pom86}, we expect to find intermittency in the context
of directed percolation. This means the occurence of short-range
self-similar fluctuations
which dominate the average characteristics of a multiparticle system.

This rough model description motivates our simulations of $1+1$ dimensional
directed percolation for particle physics. However it is worth noting that
$2+1$-dimensional directed percolation has another more theoretical
relation with multi-particle collisions, thanks to its continuum limit
in terms of the Reggeon Field Theory. In this context, the "time"
dimension corresponds to the rapidity of produced particles in a very energetic
collision, and the two-dimensional "space" is the transverse impact parameter.
So, some of our results can find a prolongation in a 2+1-dimensional
study of direct relevance to the theory of particles.

\subsection{Factorial moments}

The factorial moment analysis \cite{Bia86} is useful to avoid
as much as possible a "contamination by statistical noise" of the dynamical
fluctuations one is looking for. The general idea is
that a large part
of the non-dynamical statistical fluctuations is due to the small number
of particules (statistical realizations), and thus can be assimilated to
a "noise" with simple stochastic characteristics. In the absence of other
information, the "noise" is assumed to be Poissonian and thus can be
de-convoluted from
the observed distribution of particles by the factorial moments' trick.
Of course, this assumption looks strong, but many model tests
\cite{Bia86,Bia91} have shown its validity in multi-particle physics. On the
other hand,
statistical systems like directed percolation are useful to check the
"noise" assumption in cases where it is embedded in the dynamics, and allow
to deepen the statistical analysis.

In this specific case, however, one disposes of an information on the
total number of sites under study in a given phase-space box, and thus,
adding this constraint to the "noise" definition, one is led to consider
the Bernoulli "noise". More specifically for lattice systems, one
writes :

\BEA
F_{q} &=& \frac{<\rho^{q}>}{<\rho>^{q}} \nonumber \\
<\rho^{q}> &=& \frac{<n(n-1)\cdots (n-q+1)>}{<N(N-1)\cdots (N-q-1)>}
\EEA
where $N$ stands for the number of sites per bin, $\rho$ is the density of
occupied sites and the average $<>$ is meant over the total number of bins.
Notice that the
choosen normalization for factorial moments has been designed for the Bernoulli
noise \cite{Bia86} expected in lattice simulations \cite{Nov89} and checked
in detailed calculations on the Ising model \cite{Gup91}. If the factorial
analysis works well, the density $\rho$ and its fluctuations are free
from statistical perturbations and the intermittency parameters $\phi_q$
and $\bar{\cal D}_q$ (see (1.2), (1.3)) can be estimated.

Intermittency studies for statistical systems using factorial moments (2.4)
are particularly interesting since they offer an unique opportunity to
discuss a scaling behaviour of fluctuations in a well-defined dynamical
scheme. There is no need of a small coupling hypothesis nor of an assumption
on the statistical noise. Indeed, the statistical fluctuations from site
to site are, together with the dynamical ones, encoded in the generic rules
(2.1).

\section{Numerical simulation}

We now turn to describe our simulation procedure and the numerical results
for the intermittency indices.
We use directed bond percolation on a square
lattice. Since an intermittent behaviour as the one in
eq.~(1.2) can only occur at criticality, we shall always work
at the critical point $p=p_c$, see (2.2). As the initial state, we take
occupied sites $(1)$ on all
spatial sites at time $t=0$. Both periodic and free boundary
conditions will be considered, see below.

The factorial moments can be calculated for a two-dimensional or a
one-dimensional binning. In the latter case, we can either choose a
spatial (vertical) or a temporal (horizontal) one. For the calculation,
one can either take subdivisions of the entire lattice with $N$
spatial sites into account or only apply the
procedure of repeated subdivions to a certain subcell of $L < N$ spatial
sites. Both of these
procedures were tried for the case of periodic boundary conditions, while
for free boundary conditions, only the second one was used. The moments
were obtained calculating by generating $L$ time steps of the evolution
of the directed percolation cluster. The resulting square of $L \times L$ sites
was then subdivided into squares of sizes $L /2^{b} \times L/2^{b}$ with
$b=0,1,2,\ldots$ to obtain the two-dimensional factorial moments
$F_{q}^{(2D)}$.
To obtain the one-dimensional moments $F_{q}^{(H)}$ and $F_{q}^{(V)}$,
in the same square for the fixed time (for $F_{q}^{(H)}$) or for fixed
position (for $F_{q}^{(V)}$) subcells of length $L/2^{b}$ were formed, see
fig.~2.
We take an increment of five "layers" (intermediate steps in time or space,
respectively) between two consecutive calculations
for the $1D$ factorial moments with a given square of size $L\times L$.
To reduce the statistical correlations between measurements, we generated
2000 "thermalisation" time evolutions steps before beginning the calculation
of the factorial moments and generate an additional 2000 time steps between
any two squares to be considered for measurements. Also, if the
directed percolation cluster happened to die out (this only occured for
fairly small lattices), we reinitialized with
all spatial sites occupied and 2000 thermalisation steps.
Since correlation function
studies have shown that scaling sets in after about 20 thermalization steps
(see \cite{Ben84}), this thermalization should be sufficient. Almost the
entire CPU time was spent in calculating the factorial moments.

We now discuss the necessary spatial lattice size in order to become
independent of finite-size effects. To see this, we examine
the calculated factorial moments $F_{4}(\delta)$ as function of the
binning size $\delta$. In fig.~3-a, we display the Renyi codimension
$\bar{\cal D}_{4}$ as calculated for boxes of linear size $L/2^{b}$ for
the $2D$ factorial moment $F_{4}^{(2D)}$. Similarly, fig.~3-b shows the
same quantity as obtained from $F_{4}^{(H)}$ and fig.~3-c those obtained
from $F_{4}^{(V)}$.

We first observe that for all three cases the apparent Renyi codimensions
become independent of the lattice size $N$, the size $L$ of the
largest box considered and of the boundary conditions if $N$ is larger
than 4096. These sizes are somewhat larger than the largest ones considered
in intermittency model calculations so far (see e.g. \cite{Gup91}).
We can conclude that the data for $N=8192$ will describe accurately
the behaviour of the $N\rar\infty$ limit and that finite-size effects
are small for this size. Secondly, we see that, apart from the largest
blocks (where $b$ is small) and the smallest one (where the cells just
contains two sites), the data show a large and wide plateau and become
$b$- independent. The value of the plateau will be taken as our estimate
for the Renyi codimensions. These observations hold for all the three types
of factorial moments considered. We have drawn similar plots for the
cases $q=2$ and $q=6$ and find the qualitative behaviour to be the same,
although the convergence towards the $N\rar\infty$ limit is somewhat slower
for higher rank.
We can conclude that our directed bond percolation model at $p=p_c$ does indeed
show intermittent behaviour. For a precise calculation of the intermittency
indices, we have seen that lattices $N=8192$ spatial sites are
large enough. We have also studied the effect of varying $p$ away from
$p_c$. Already for variations of order $\delta p \simeq 10^{-2}$,
the $\bar{\cal D}_{q}$ are no longer almost
independent of $b$, but are rather seen to rise steeply with
increasing values of $b$.

To investigate the $q$- dependence, we take $N=8192$, $L=2048$ with
periodic boundary conditions. We generated 500 squares of size
$L \times L$ for the calculation of the $F_{q}$, with 2000 intermediate
time steps in between each square. In fig.~4-a, we give the curves
for the Renyi codimension
$\bar{\cal D}_{q}^{(2D)}$, for $q=2,3,4,5,6$ as calculated
from $F_{q}^{(2D)}$. Similarly, fig.~4-b shows the results obtained from
$F_{q}^{(H)}$ and fig.~4-c those from $F_{q}^{(V)}$. Already a first glance
at the figures shows that the $2D$ and the horizontal $(H)$ results are very
similar to each other while the vertical $(V)$ ones are quite distinct.
In both figs.~4-a and 4-b, we see that for a large range of binning
boxes (as given by $b$), the apparent Renyi codimension $\bar{\cal D}_{q}$, is
in fact $q$- independent. This means that the $2D$ and the horizontal
intermittency behaviour is in fact described by a simple fractal structure.
We read off the final Renyi codimensions
\BEQ
\bar{\cal D}^{(2D)} = 0.023 \pm 0.004
\EEQ
for the $2D$ factorial moments and
\BEQ
\bar{\cal D}^{(H)} = 0.021 \pm 0.004
\EEQ
for the horizontal factorial moments. In section~4, we shall return to
an interpretation of these results.

On the other hand, the data obtained from the vertical factorial moments
show quite a distinct behaviour, see fig.~4-c. It is not possible to
define anything like
a single Renyi codimension. To our best knowlegde, this is the first example
of a system displaying such a strong anisotropy in the factorial
moments. We note that for the maximal values of $b$, the effective
$\bar{\cal D}_{q}$
are close to the values obtained from the $2D$ or the horizontal
case. This is to be expected since for these very small boxes the effects
of the anisotropy of the directed percolation clusters have not yet
become apparent.

\section{Scaling relations and intermittency indices}

The features revealed by our numerical simulations and the estimates
(3.1) and (3.2) are quite unexpected and thus require some theoretical
interpretation. In particular, the obtained intermittency dimensions
are much smaller (while non-zero) than the exponent ratios
$\beta /\nu_{\|}$, $\beta/\nu_{\perp}$, (see (2.3)), contrary to some
expectation in non-directed percolation \cite{Sat89}. Moreover, the similarity
between $\bar{\cal D}^{(2D)}$ and $\bar{\cal D}^{(H)}$ has to be explained.
We shall now do so by using a simple scaling argument, eventually arriving
at eq.~(1.4).

Let us first consider the two-dimensional second moment:
$$F_{2}^{(2D)} := <n(n-1)>/<n>^{2}$$ where one keeps the simple normalization
(1.1) which is sufficient for the theoretical discussion in the continuum
limit. As already noticed in
the literature \cite{Bia91,Sat89}, the intermittency behaviour (1.2) is to be
related to an
effective singularity in the relevant correlation function. In the case of the
2-D moment, one may write:
\BEQ
<n(n-1)> \mathrel {\mathop {\propto } \limits_{\delta \rightarrow 0}}
 \int_{[\delta ]} dt_1 dt_2 dx_1 dx_2 \, G^{(2)} (x_1,x_2;t_1,t_2)
\EEQ
where $G^{(2)}$ is the two-point correlation function and
the integral is over the phase-space box of unit length $\delta $, see
fig.~2. This corresponds to the probability of a connecting path of occupied
sites between the space-time points
$(x_1 ,t_1 )$ and $(x_2 ,t_2)$. Note an implicit averaging over phase-space
boxes which is allowed by translational invariance, well verified in our
simulations. It is also important to notice that, for practical simulation
purpose, the binsize $\delta $ has to be considered very small, but large
with respect to the bond size.

The scaling behaviour of $G^{(2)}$ near the percolation transition point
$p_{c}$
can be written:
\BEQ
G^{(2)} \sim {(\delta p)}^{2\beta } \Psi[{(\delta p)}^{{\nu }_{\perp }}\theta
 ;{(\delta
p)}^{{\nu }_{||}}\tau ] \sim {\theta}^{-\frac{2\beta}{{\nu }_{\perp}}}
\cdot ~\mbox{\rm cste}
\EEQ
where $\delta p = p-p_{c}$, $\theta = |x_1 - x_2|$ and $\tau = |t_1 - t_2|$ and
$\Psi$ is a scaling function.

The first part of relation (4.2) comes from the critical behaviour at the
percolation transition while the second part is the
dominant $\delta p$-independent
behaviour after rescaling. Note that the known strong anisotropy in time
of directed percolation provides a hint for the dominance of the $\theta$
singularity over the phase-space box. This typical feature of directed
percolation is well demonstrated by the narrowness of percolation clusters in
 the "time" direction, see ref. \cite{Hed91}.Inserting the behaviour (4.2)
in the estimate (4.1) at small $\delta $ one finds:
\BEQ
<n(n-1)> \sim {\delta }^{4-2\beta /{\nu }_{\perp }}
\EEQ
the exponent $4$ being simply interpreted as the naive integration dimension in
(4.1).

The key remark of our analysis now coming is related to the normalization
denominator ${<n>}^2$ of
$F_{2}^{(2D)}$. An oversimplified evaluation of this denominator would have led
to some number proportional to the phase-space, namely $\delta ^4$, and
thus to an intermittency exponent $\beta /\nu _{\perp }$. This is much too
large as compared
with the numerical value (3.1). Rather, one has to take into account
the density of occupied sites in the box $[\delta ]$, which is not uniform.
This density  is better described by a fractal set whose
dimension $D$ has been recently
numerically estimated from the formation of very long percolation
clusters \cite {Hed91}. Using this information, we write:
\BEQ
<n> \sim \delta^{D} = \delta^{2-\bar{D}}
\EEQ
introducing the fractal dimension of occupied sites and its
codimension $\bar{D}$. We thus get the general relation:
\BEQ
F_{2}^{(2D)} \sim\ {(\delta^{-2})}^{(\beta /\nu _{\perp }-{\bar{D})}}
\EEQ
Comparing with (1.3), we note that since the boxes considered are
two-dimensional, we have $F_{q}^{(2D)}\sim (\delta^{-2})^{\phi_q}$.
In terms of the Renyi codimensions $\rcd{q}^{(2D)}$ of a (fractal) intermittent
process, we recover indeed the relation (1.4). The factorial moment satisfies
\BEQ
F_{2}^{(2D)}:= \frac{<n(n-1)>}{<n>^{2}} \sim \frac{<\rho ^{2}>}{<\rho >^{2}}
\ge  1
\EEQ
which implies in turn that the Renyi codimension $\bar{\cal D}$ is
non-negative. From our numerical estimate (3.1), we see that
the fractal cluster codimension $\bar{D}$ and the exponent ratio
$\beta/\nu_{\perp}$ are very close, up to a small difference of definite sign.

It is of interest to compare this result with earlier determinations of the
fractal dimension of directed percolation clusters.
A careful study of this question was presented in \cite{Hed91}. Three
possibilities to calculate $\bar{D}$ were considered. The first one
simply assumes that the two-dimensional fractal can be considered as
a direct product. The second possibility relies on a deterministic
growth argument. Finally, the third possibility proceeds along a scaling
argument taking into account the presence of the two length scales given
by $\xi_{\|}$ and $\xi_{\perp}$ \cite{Nad84}. These arguments lead to
\cite{Hed91}
\BEQ
\bar{D}^{(2D)} = \left\{ \begin{array}{rl}
\beta\left( \frac{1}{\nu_{\|}} + \frac{1}{\nu_{\perp}} \right) \simeq 0.412~ &
{}~\mbox{\rm direct product} \\[1.4\baselineskip]
2\beta/\left(\nu_{\|} + \nu_{\perp}\right) \simeq 0.196~ & ~\mbox{\rm
 deterministic
growth} \\[1.4\baselineskip]
\beta/\nu_{\perp} \simeq 0.253~ & ~\mbox{scaling \cite{Nad84}}
\end{array} \right. \label{FracDim}
\EEQ
Via (1.4), this would lead to $\bar{\cal D}^{(2D)} \simeq -0.16, 0.06$ and $0$,
respectively. A precise numerical calculation using box-counting arguments
yields \cite{Hed91}
\BEQ
D = 1.765 \pm 0.005
\EEQ
and one notes a discrepancy with all of the relations (\ref{FracDim}).

However, (4.8) would yield via (1.4) a value $\bar{\cal D}^{(2D)} = 0.018(5)$
in agreement with our estimate (3.1). In other terms, one may
interpret the small but non-zero difference between
the cluster and correlation fractal dimensions as due to a weak intermittency
phenomenon in $1+1$-dimensional directed percolation.

 The same relation (1.4) holds for the "space" moments $F_{2}^{(H)}$, but
 it requires a more refined way of derivation, since the 2-point
 correlation function is zero between 2 sites with no "time" separation,
 see fig.~2. In this particular case, one has to consider a 3-point
 correlation function $G^{(3)}$ linking the two given sites $(x_{1},t_{1})$ and
 $(x_{2},t_{2})$, starting from an initial  $(x_{0},t_{0})$. One would
 assume the scaling behaviour

\BEQ
G^{(3)} \sim {(\delta p)}^{3\beta } \Phi\left[
(\delta p)^{\nu_{\perp}}\theta_{ij} ;
(\delta p)^{\nu_{\|}}\tau_{ij} \right]
\sim \theta_{12}^{-\frac{3\beta}{\nu_{\perp}}} \cdot ~\mbox{\rm cste}
\EEQ
where $\delta p = p-p_{c}$, $\theta_{ij} = |x_i - x_j|$ and
$\tau_{ij} = |t_i - t_j|$ and
$i,j = 0,1,2; i\ne j$. Integrating over the initial conditions  at
 $(x_{0},t_{0})$, and the phase-space box of linear dimension $\delta$ for
  $(x_{1},t_{1})$ and
 $(x_{2},t_{1})$ (note that we are at equal time), one gets:
\BEA
<n(n-1)>    & \sim & \delta^{D_{\perp } +1 -\beta / \nu _{\perp }} \nonumber \\
<n>        & \sim & \delta^{D_{\perp }} \nonumber \\
F_{2}^{(H)}  & \sim & \delta ^{-(\beta / \nu _{\perp} - {\bar {D}}_{\perp })}
\EEA
where $D_{\perp}$ is the "horizontal" fractal dimension.
One finds again the same type of relation (1.4), which thus appears
quite general.

We also note that our numerical estimates (3.1) and (3.2) suggest that the
fractal codimensions $\bar{D}^{(2D)} \simeq \bar{D}_{\perp}$, which is in
agreement with the literature \cite{Nad84,Hed91}.

A simple relation as (1.4) would have to be modified in the case of the "time"
dimension, due to the multifractal character manifested by our
numerical simulation. It is to be noticed that such a feature is
in general associated with non-equilibrium cascading processes \cite{Pes89}.
It is thus to our knowledge, the first time that multifractal
fluctuations appear in a percolation problem, in probable relation with
the oriented "time" direction characteristic of directed percolation.

\section{Conclusions and outlook}

Let us summarize our results:

\begin{enumerate}
\item On the basis of a simulation of directed bond percolation
in 1-"space"-1-"time" dimensions and after checking against statistical
and boundary effects the stability of our results, we have found that
intermittency patterns of fluctuations show up in the distribution
of occupied sites clusters. The fluctuations along the "space" dimension
and in two dimensions are dominated by weak, fractal, structures of
Renyi codimension
$\bar {\cal D} = 0.022 \pm 0.004$, (the mean value of (3.1) and (3.2))
while "time"-like fluctuations appear
with a more complicated structure of multi-fractal type.

\item These numerical results can be interpreted as resulting from the
scaling behaviour of the two- and three-point correlation function
in the continuum limit at the bond percolation critical point, both
corrected by  the fractal dimension of clusters.
The overall result is summarized in formula (1.4), which appears
well verified for the two recognized fractal cases, "spatial" and
two-dimensional. These two cases were seen to have the same intermittency
indices. Our result clarifies a discussion on the calculation of
fractal dimensions of directed percolation in the literature \cite{Hed91}.
\end{enumerate}

It is quite interesting to note that, in its generalized form (1.4), the
relation expressing
the intermittency indices (in the fractal case) as a function of the
correlation indices and the fractal cluster dimension can be applied to other
problems. Let us mention the case of percolation at a second-order phase
transition \cite{Wos88}, which is also fractal, according to general scaling
arguments \cite{Sat89}. An interesting example is the two-dimensional
Ising spin system, where the relation (1.4) leads to the following
prediction:

\BEQ
\bar{\cal D} = \beta /\nu  - \bar{D} = 7/96
\EEQ
where the exponent ratio $\beta/\nu =1/8$ is well known and
the value of the fractal codimension $\bar{D}=5/96$ can be obtained by
conformal invariance techniques \cite{Dup88}.
The result (5.1), which is in agreement with numerical simulations
\cite{Gup91}, and finite-size scaling evaluations \cite{Bur92}, clarifies
a set of contradicting issues on the subject. While different from the
value $\bar{\cal D} = \beta /\nu$, predicted in ref. \cite{Sat89},
the value (5.1) is different from the fractal dimension $\bar{D}$
of clusters as well. Indeed, the system admits
a set of {\it fractal} fluctuations around a {\it fractal} object, which here
was the average percolation cluster. We also remark that the value found
in \cite{Bur92} is not accidental, as conjectured by the authors, but
a prediction valid at the continuum limit.

Finally, the access to so refined quantities as fluctuation moments
allowed by numerical calculations of directed percolation is an
incentive to perform the same analysis in 2+1 dimensions, with the
bonus of a direct application to particle physics through the Reggeon Field
model. Note that the already obtained result in 1+1 dimension, applied
to the Swinger-type model \cite{Bia89,Bar87}, confirm the existence of
weak intermittent fluctuations, seen in the model simulations \cite{Bia89}.

\zeile{3}
{\bf Acknowledgements}
\zeile{2}
We acknowledge R. Bidaux for a valuable discussion and Y. Leroyer for a
careful reading of the manuscript. The present work
was initiated at CERN (Theory Division) and Geneva University.
One of us (MH) has the pleasure to thank the Service de Physique
Th\'eorique at CEA Saclay for its warm hospitality while this
work was finished.

\newpage
{\bf Figure captions}
\zeile{3}\noindent
Figure~1: The evolution as described by directed bond percolation.
The result of a simulation on a small lattice with 32 "spatial" sites and
periodic boundary conditions is shown. The triangles
show some typical evolution cells whose conditional probabilities are recalled
below the simulation sample. Note that, by convention, white dots are for empty
sites and stars for occupied ones.
\zeile{1}\noindent
Figure~2: Binning for the definition of the factorial moments. Typical binning
boxes are represented on the simulated patterns of sites (same series as
Fig.1): Two-dimensional, Horizontal and Vertical examples are shown, together
with typical connected paths which contribute to the singular behaviour at the
continuum  limit (see text).
\zeile{1}\noindent
Figure~3: Effective Renyi codimension $\rcd{2}$ as obtained from the
factorial moment (a) $F_{4}^{(2D)}$, (b) $F_{4}^{(H)}$, (c)$F_{4}^{(V)}$
as function of the binning size $L/2^{b}$. The curves corresponds to the
following lattices $\bigcirc :~ N=8192, L=2048$ periodic boundary conditions
(full curve), $\bigcirc :~ N=8192, L=2048$ free boundary conditions (dotted),
$\Diamond :~ N=4096, L=1024$ free boundary conditions (dashed) and
$\Box :~ N=4096, L=4096$ periodic boundary conditions (dash-dotted). Notice
the quite larger vertical coordinate  scale in fig.(c).
\zeile{1}\noindent
Figure~4: Renyi codimensions $\rcd{q}= \phi_q /(q-1)$ as
obtained for $N=8192$ sites,
a maximal box size of $L=2048$ and periodic boundary conditions
as obtained from the factorial moments (a) $F_{q}^{(2D)}$,
(b) $F_{q}^{(H)}$, (c) $F_{q}^{(V)}$. The symbols correspond to the following
values of q $\bigcirc ~q=2$, $\bigcirc ~q=3$, $\Diamond ~q=4$,
$\Box ~q=5$ and $\Box ~q=6$. Notice
the quite larger vertical  coordinate   scale in fig.(c).


\end{document}